\documentclass[aps,prl,superscriptaddress,twocolumn,nofootinbib]{revtex4}
\usepackage[dvips]{graphicx}
\usepackage{bm,latexsym,amsmath,amssymb,amsfonts,mathrsfs}
\usepackage{color}
\input{colordvi.tex}

\begin{document}

\title{Non-Gaussianities of primordial perturbations and tensor sound
speed}

\author{Toshifumi~Noumi}
\email[Email: ]{toshifumi.noumi"at"riken.jp}
\affiliation{Mathematical Physics Laboratory, RIKEN Nishina Center,
Saitama 351-0198, JAPAN}

\author{Masahide~Yamaguchi}
\email[Email: ]{gucci"at"phys.titech.ac.jp}
\affiliation{Department of Physics, Tokyo Institute of Technology, Tokyo
152-8551, Japan}

\begin{abstract}

We investigate the relation between the non-Gaussianities of the
primordial perturbations and the sound speed of the tensor
perturbations, that is, the propagation speed of the gravitational
waves. We find that the sound speed of the tensor perturbations is
directly related not to the auto-bispectrum of the tensor perturbations
but to the cross-bispectrum of the primordial perturbations, especially,
the scalar-tensor-tensor bispectrum. This result is in sharp contrast
with the case of the scalar (curvature) perturbations, where their
reduced sound speed enhances their auto-bispectrum. Our findings
indicate that the scalar-tensor-tensor bispectrum can be a powerful tool
to probe the sound speed of the tensor perturbations.
\end{abstract}

\preprint{RIKEN-MP-85}
\maketitle

\subsection{Introduction}

Inflation is now widely accepted as a paradigm of early Universe to
explain the origin of the primordial perturbations as well as to solve
the horizon and the flatness problems of the standard big-bang cosmology
\cite{inflation}. The current observational data such as the cosmic
microwave background (CMB) anisotropies support almost scale-invariant,
adiabatic, and Gaussian primordial curvature fluctuations as predicted by
inflation.
While the paradigm itself is well established and widely
accepted, its detailed dynamics, e.g. the identification of an inflaton,
its kinetic and potential structure,
and its gravitational coupling,
are still unknown.

The non-Gaussianities of primordial curvature perturbations are
powerful tools to give such informations. It is well-known that the
equilateral type of bispectrum of primordial curvature perturbations is
enhanced by the inverse of their sound speed squared
\cite{Seery:2005wm,Chen:2006nt}. The null observation of the equilateral
type by the Planck satellite, characterized as $f^{\rm equil}_{\rm NL} =
42 \pm 75$ (68\% CL)~\cite{Ade:2013ydc}, yields stringent constraints on
the sound speed of the curvature perturbations as $c_s \ge 0.02$ (95\%
CL)~\cite{Ade:2013ydc}.
The local type of bispectrum of primordial
curvature perturbations also gives useful informations.
Maldacena's
consistency relation \cite{Maldacena:2002vr,Creminelli:2004yq} says that
the parameter $f_{\rm NL}^{\rm local}$ characterizing the local type of
bispectrum is as much as ${\cal O}(0.01)$ for a single field inflation
model.\footnote{This consistency relation is derived under some
reasonable assumptions. If we violate some of them, there is a
counterexample, which is given in Ref.~\cite{Namjoo:2012aa} for
example.} Though the current constraint on this parameter by the Planck
satellite is already tight as $f_{\rm NL}^{\rm local} = 2.7 \pm 5.8$
(68\% CL)~\cite{Ade:2013ydc}, the detection of non-Gaussianities
with $f_{\rm NL}^{\rm local} \sim {\cal O}(1)$ by future observations
would rule out a single field inflation model.

Inflation generates not only primordial curvature perturbations but also
primordial tensor perturbations \cite{Starobinsky:1979ty}. Very
recently, it was reported that primordial tensor perturbations have been
found and the tensor-to-scalar ratio $r$ is given by $r =
0.20^{+0.07}_{-0.05}$ (68\% CL)~\cite{Ade:2014xna}, though it is
constrained as $r < 0.11$(95\% CL) in the Planck results
\cite{Ade:2013uln}. Their amplitude directly determines the energy scale
of inflation, so it is estimated as $V^{1/4} \simeq 2.2 \times 10^{16}$
GeV given $r \simeq 0.2$ \cite{Ade:2014xna} and $P_{\zeta} \simeq 2.2
\times 10^{-9}$ \cite{Ade:2013uln}.  If we go beyond the powerspectrum,
it is known that the bispectra of primordial tensor perturbations enable
us to probe the gravitational coupling of the inflaton field
\cite{Gao:2011vs}. Such a non-trivial gravitational coupling easily
modifies the sound speed of primordial tensor perturbations,
$c_{\gamma}$, and it can significantly deviate from unity
\cite{Kobayashi:2011nu}. Then, one may wonder if the small sound
velocity of primordial tensor perturbations can enhance their
non-Gaussianities as in the case of the curvature perturbations. In this
{\it Letter}, we are going to address this issue.

The relation between the sound speed and the non-Gaussianities of
primordial curvature perturbations can be clearly understood by use of
the effective field theory (EFT) approach to
inflation~\cite{Cheung:2007st}. Inflation can be characterized by the
breakdown of time-diffeomorphism invariance due to the time-dependent
cosmological background and the general action for inflation can be
constructed based on this symmetry breaking structure. 
The primordial
curvature perturbation can be identified with the Goldstone mode $\pi$
associated with the breaking of time-diffeomorphism invariance.
The primordial
curvature perturbation can be identified with the Goldstone mode $\pi$
associated with the breaking of time-diffeomorphism invariance.
The symmetry arguments require that modification of the sound speed $c_s$
induces non-negligible cubic interactions of the Goldstone mode $\pi$,
and hence
the sound speed and the bispectrum of the curvature perturbations are directly related.

In this {\it Letter}, we investigate the relation between the sound
speed of tensor perturbations and the bispectrum of primordial
perturbations, based on the EFT approach.  We first identify what kind
of operators can modify the tensor sound speed.  Then, we clarify which
type of bispectrum arises associated with the modification and can be
used as a probe for the tensor sound speed.

\subsection{The EFT approach}

We start from a brief review of the EFT approach~\cite{Cheung:2007st}
and clarify our setup.  In the unitary gauge, where the inflaton field
does not fluctuate, dynamical degrees of freedom in single-clock
inflation are the metric field $g_{\mu\nu}$ only and the action should
respect the (time-dependent) spatial diffeomorphism invariance.  The
action at the lowest order in perturbations can be uniquely determined
by the background equations of motion and the residual spatial
diffeomorphism invariance as
\begin{align}
\label{simplest_action}
S_{0}&=M_{\rm Pl}^2\int d^4x \sqrt{-g}\left[
\frac{1}{2}R
+\dot{H}g^{00}
-(3H^2+\dot{H})\right]\,,
\end{align}
where $H(t)=\dot{a}/a$ is the background Hubble parameter
with $a(t)$ being the the background scale factor
\begin{align}
ds^2=-dt^2+a(t)^2d\vec{x}^2\,.
\end{align}
This simplest action describes tree-level dynamics of the
single-field inflation with a canonical kinetic term in the Einstein
gravity.  Modifications of inflation models and quantum corrections can
be described by including higher order perturbation terms.  Ingredients
for higher order perturbations are $\delta g^{00}$, $\delta K_{\mu\nu}$,
$\delta R_{\mu\nu\rho\sigma}$, and their derivatives:
\begin{align}
\nonumber
&\delta g^{00}=g^{00}+1\,,
\quad
\delta K_{\mu\nu}=K_{\mu\nu}-Hh_{\mu\nu}\,,\\
\nonumber
&\delta R_{\mu\nu\rho\sigma}=
R_{\mu\nu\rho\sigma}-H^2(h_{\mu\rho}h_{\nu\sigma}-h_{\mu\sigma}h_{\nu\rho})\\
&\qquad\qquad\,\,
+(H^2+\dot{H})(h_{\mu\rho}n_\nu n_\sigma+\text{$3$ terms})\,,
\end{align}
which are covariant under the spatial diffeomorphism and vanish on the
FRW background.  Here $n_\mu=-\frac{\delta^0_\mu}{\sqrt{-g^{00}}}$ is
the unit vector perpendicular to constant-$t$ surfaces,
$h_{\mu\nu}=g_{\mu\nu}+n_\mu n_\nu$ is the induced spatial metric, and
$K_{\mu\nu}=h_\mu^\sigma\nabla_\sigma n_\nu$ is the extrinsic curvature
on the spatial slices.  We define the scalar curvature perturbation
$\zeta$ and the tensor perturbation $\gamma_{ij}$ as
\begin{align}
h_{ij}=a^2e^{2\zeta}(e^\gamma)_{ij}
\quad
{\rm with}
\quad
\gamma_{ii}=\partial_i\gamma_{ij}=0\,.
\end{align}
The general action for single-clock inflation
can then be expanded in perturbations and derivatives as~\cite{Cheung:2007st}
\begin{align}
\nonumber
S&=S_0+\int d^4x \sqrt{-g}\left[
\frac{M_2(t)^4}{2}(\delta g^{00})^2
-\frac{\bar{M}_1(t)^3}{2}\delta g^{00}\delta K\right.\\
\label{EFT_action}
&\qquad\,\,
\left.
-\frac{\bar{M}_2(t)^2}{2}\delta K^2
-\frac{\bar{M}_3(t)^2}{2}\delta K_\mu^\nu\delta K_\nu^\mu
+\ldots
\right]\,,
\end{align}
where $\delta K=\delta K_\mu^\mu$ and the dots stand for higher
derivative terms and higher order perturbations.  The above four
correction terms are the only operators relevant
to the dispersion relations of primordial perturbations
in the decoupling limit,
with up to two derivatives on metric
perturbations, and without higher time derivatives such
as~$\ddot{\zeta}$. In the following we focus on these operators
(see~\cite{full} for more general cases).

\subsection{Tensor sound speed and the powerspectrum}

We now investigate the tensor perturbations $\gamma_{ij}$ based on the
EFT framework.  Among the operators displayed in~(\ref{EFT_action}),
only $\delta K_\mu^\nu\delta K_\nu^\mu$
except the Einstein-Hilbert
action induces the second order tensor perturbations:
\begin{align}
\label{tensor_sound_correction}
\delta K_\mu^\nu\delta K_\nu^\mu
\ni -\frac{1}{4}(\partial_te^{-\gamma}\partial_te^\gamma)_{ii}=\frac{1}{4}(\dot{\gamma}_{ij})^2
+\mathcal{O}(\gamma^4)\,,
\end{align}
which deforms the kinetic term for $\gamma$ as
\begin{align}
\int d^4x\,a^3
\frac{M_{\rm Pl}^2}{8}
c_{\gamma}^{-2}
\left[(\dot{\gamma}_{ij})^2-c_{\gamma}^{2}\frac{(\partial_k\gamma_{ij})^2}{a^2}
\right]\,.
\end{align}
Here the tensor sound speed $c_\gamma$ is given by
\begin{align}
c_\gamma^2
=\frac{M_{\rm Pl}^2}{M_{\rm Pl}^2-\bar{M}_3^2}\,.
\end{align}
To compute the powerspectrum, let us decompose $\gamma_{ij}$ into the
two helicity modes as
\begin{align}
\gamma_{ij}&=\int\frac{d^3k}{(2\pi)^3}
\sum_{s=\pm}\epsilon_{ij}^s({\bf k})\gamma_{\bf k}^s(t)e^{i{\bf k}\cdot {\bf x}}\,,
\end{align}
where $s=\pm$ is the helicity index.
The polarization tensor $\epsilon_{ij}^s({\bf k})$
is symmetric, traceless, and transverse.
Its normalization and reality conditions can be stated as
\begin{align}
\sum_{i,j}\epsilon_{ij}^s({\bf k})\epsilon_{ij}^{s^\prime}(-{\bf k})=2\delta_{ss^\prime}\,,
\quad
\left(\epsilon_{ij}^{s}({\bf k})\right)^\ast=\epsilon_{ij}^{s}(-{\bf k})\,.
\end{align}
These two helicity modes are quantized as
\begin{align}
\gamma_{\bf k}^s(t)=b_{s,{\bf k}}v_k(t)+b_{s,-{\bf k}}^\dagger v^*_k(t)
\end{align}
with the standard commutation relation
\begin{align}
[b_{s,{\bf k}},b^\dagger_{s',{\bf k}'}]=\delta_{ss'}(2\pi)^3\delta({\bf k}-{\bf k}^\prime).
\end{align}
Here and in what follows, we neglect the time-dependence of the Hubble
parameter $H$ and the sound speed $c_\gamma$.  The mode function $v_k$
for the Bunch-Davies vacuum is then given~by
\begin{align}
v_k=\frac{H}{M_{\rm Pl}}\frac{c_\gamma}{(c_\gamma k)^{3/2}}
(1+ic_\gamma k\tau)e^{-ic_\gamma k\tau}\,,
\end{align}
where $\tau$ is the conformal time $ad\tau=dt$.
With this mode function,
the tensor two point function is calculated as
\begin{align}
\nonumber
\langle\gamma_{{\bf k}}^s\gamma_{{\bf k}^\prime}^{s^\prime}\rangle
=\delta_{ss^\prime}(2\pi)^3\delta^{(3)}({\bf k}-{\bf k}^\prime)\frac{\pi^2}{2k^3}\mathcal{P}_\gamma(k)
\\
{\rm with}
\quad
\mathcal{P}_\gamma(k)=c_\gamma^{-1}\cdot\frac{2H^2}{\pi^2M_{\rm Pl}^2}\,.
\end{align}
Note that the tensor powerspectrum $\mathcal{P}_\gamma$ is proportional
to $c_\gamma^{-1}$, and therefore, the tensor-to-scalar ratio
$r=\mathcal{P}_\gamma/\mathcal{P}_\zeta$ has a negative correlation with
the tensor sound speed $r\propto c_\gamma^{-1}$.

\subsection{Tensor bispectrum}

Let us then discuss the relation between the tensor sound speed
$c_\gamma$ and the tensor bispectrum.  An important point is that no
tensor cubic interactions arise from the operator $\delta
K_\mu^\nu\delta K_\nu^\mu$ as shown in
(\ref{tensor_sound_correction}), in contrast to the scalar sound speed
case~\cite{Cheung:2007st}.  If we concentrate on the operators displayed
in (\ref{EFT_action}), the only source of tensor cubic interactions is
the Einstein Hilbert term in $S_0$:
\begin{align}
\label{S_0_tensor^3}
S_0\ni
M_{\rm Pl}^2\int d^4x \,\frac{a}{4}
\left(\gamma_{ik}\gamma_{jl}-\frac{1}{2}\gamma_{ij}\gamma_{kl}\right)
\partial_k\partial_l\gamma_{ij}\,.
\end{align}
The deformation of the tensor sound speed can
then affect tensor bispectra only through the change in the field
normalization and the sound horizon.

For qualitative understanding of these effects, let us first perform an
order estimation of the nonlinearity parameter.  For this purpose, it is
convenient to work in the real coordinate space, rather than in the
momentum space.  In the real coordinate space, the two-point function is
estimated as
\begin{align}
\langle\gamma\gamma\rangle
&\sim c_\gamma^2\frac{H^2}{M_{\rm Pl}^2}\,.
\end{align}
On the other hand,
the three-point function originated from the cubic interaction (\ref{S_0_tensor^3})
can be estimated as
\begin{align}
\langle\gamma\gamma\gamma\rangle
&\sim\left(\frac{M_{\rm Pl}^2}{c_\gamma^2H^2}\right)\cdot\left(c_\gamma^2\frac{H^2}{M_{\rm Pl}^2}\right)^3\,,
\end{align}
where the first factor arises from the vertex (\ref{S_0_tensor^3}) and
we used $\displaystyle\frac{\partial_i}{a}= c_\gamma^{-1}\cdot
c_\gamma\frac{\partial_i}{a}\sim c_\gamma^{-1}H$.  The second factor is
from the three tensor propagators.  We can then estimate the
nonlinearity parameter (normalized by the tensor powerspectrum) as
\begin{align}
\widetilde{f}_{{\rm NL},\gamma}\sim
\frac{\langle\gamma\gamma\gamma\rangle}{\langle\gamma\gamma\rangle^2}\sim1\,,
\end{align}
which is of the order one and independent of the tensor sound speed $c_\gamma$.

For more details, we present the result of the momentum space analysis
briefly.  Taking into account the modification of the field
normalization and the sound horizon, we can easily factorize the
$c_\gamma$-dependence of the bispectrum as
\begin{align}
\langle\gamma_{{\bf k}_1}^{s_1}\gamma_{{\bf k}_2}^{s_2}\gamma_{{\bf k}_3}^{s_3}\rangle
&=c_\gamma^{-2}\cdot\langle\gamma_{{\bf k}_1}^{s_1}\gamma_{{\bf k}_2}^{s_2}\gamma_{{\bf k}_3}^{s_3}\rangle\Big|_{S_0}\,,
\end{align}
where $\langle\gamma_{{\bf k}_1}^{s_1}\gamma_{{\bf
k}_2}^{s_2}\gamma_{{\bf k}_3}^{s_3}\rangle\Big|_{S_0}$ represents
the three-point function taking into account only $S_0$ (that is,
$c_{\gamma}=1$ case)~\cite{Maldacena:2002vr}. In terms of the shape
function $\widetilde{S}_{s_1,s_2,s_3}$ normalized by the tensor powerspectrum,
\begin{align}
\nonumber
&\langle\gamma_{{\bf k}_1}^{s_1}\gamma_{{\bf k}_2}^{s_2}\gamma_{{\bf k}_3}^{s_3}\rangle
\\
\label{ttt}
&=(2\pi)^3\delta^{(3)}\Big(\sum_i{\bf k}_i\Big)
\frac{(2\pi)^4\mathcal{P}_\gamma^2}{k_1^2k_2^2k_3^2}
\widetilde{S}_{s_1,s_2,s_3}(k_1,k_2,k_3)\,,
\end{align}
the tensor bispectrum can then be expressed as\footnote{Here and in what
follows, we drop a phase factor associated with the spin-$2$ structure
of the polarization tensor for simplicity. See~\cite{full} for details.}
\begin{align}
\nonumber
\widetilde{S}_{s_1,s_2,s_3}(k_1,k_2,k_3)
=\frac{\sqrt{2}}{8192}
\frac{F(s_1k_1,s_2k_2,s_3k_3)}{k_1^3k_2^3k_3^3}\\
\times\left[
k_t-\frac{\sum_{i>j}k_ik_j}{k_t}-\frac{k_1k_2k_3}{k_t^2}
\right]\,,
\end{align}
where $k_t=k_1+k_2+k_3$ and
$F(x,y,z)$ is given by
\begin{align}
\nonumber
&F(x,y,z)\\
&=(x+y+z)^5(x+y-z)(y+z-x)(z+x-y)\,.
\end{align}
The point is that the $c_\gamma$-dependence of the three-point function can
be absorbed into the prefactor $\mathcal{P}_\gamma^2$ in (\ref{ttt}) and
the shape function is $c_\gamma$-independent.
The nonlinearity
parameter defined by
\begin{align}
\widetilde{f}^{s_1s_2s_3}_{{\rm NL},\gamma}
=\frac{10}{9}\widetilde{S}_{s_1,s_2,s_3}(k,k,k)
\end{align}
can be also calculated as~\cite{Gao:2012ib}
\begin{align}
\label{widetilde_f_NL_gamma}
\widetilde{f}^{\pm\pm\pm}_{{\rm NL},\gamma}
=\frac{255\sqrt{2}}{4096}\,,
\quad
\widetilde{f}^{\pm\pm\mp}_{{\rm NL},\gamma}
=\frac{85\sqrt{2}}{110592}\,.
\end{align}
As we already discussed in the qualitative estimation,
$\widetilde{f}_{{\rm NL},\gamma}$ does not depend on the tensor sound
speed $c_\gamma$ and is of the order one.  Also note that the
nonlinearity parameter $f_{{\rm NL},\gamma}$ normalized by the scalar
powerspectrum is given by
\begin{align}
\label{f_NL_gamma}
f_{{\rm NL},\gamma}=\frac{\mathcal{P}_\gamma^2}{\mathcal{P}_\zeta^2}\,\widetilde{f}_{{\rm NL},\gamma}=r^2\, \widetilde{f}_{{\rm NL},\gamma}\sim r^2\,.
\end{align}
To summarize, the relations \eqref{widetilde_f_NL_gamma} and
\eqref{f_NL_gamma} for the nonlinearity parameters do not depend on the
sound speed $c_\gamma$ explicitly.  The shape of bispectra is also
independent of the tensor sound speed essentially because the operator
$\delta K_\mu^\nu\delta K_\nu^\mu$ does not induce tensor cubic
interactions.  In this sense, we cannot identify the tensor sound speed
only from the tensor bispectrum.  In particular, a large
$\widetilde{f}_{{\rm NL},\gamma}$ cannot be obtained unless other
operators of higher dimension or higher order perturbations are
relevant.

\subsection{Importance of cross correlations}

As we have discussed, it is not possible to determine the tensor sound
speed only from the tensor powerspectrum and bispectrum.  We now show
that cross correlations can be a useful probe for the tensor sound
speed.  For this purpose, let us perform the St\"{u}ckelberg method and
introduce the Goldstone boson $\pi$ associated with the breaking of time
diffeomorphism invariance.  By a field-dependent coordinate
transformation
\begin{align}
(t,x^i)\to(\tilde{t},\tilde{x}^i)
\quad
{\rm with}
\quad
\tilde{t}+\tilde{\pi}(\tilde{t},\tilde{x})=t\,,
\,\,
\tilde{x}^i=x^i\,,
\end{align}
the minimal action $S_0$ is transformed as
\begin{align}
\nonumber
S_0&=M_{\rm Pl}^2\int d^4x\,a^3
\bigg[
-\dot{H}\Big(
\dot{\pi}^2-\frac{(\partial_i\pi)^2}{a^2}
+\frac{\gamma_{ij}\partial_i\pi\partial_j\pi}{a^2}
\Big)
\\
&
+\frac{1}{8}\Big(
\dot{\gamma}_{ij}^2-\frac{(\partial_k\gamma_{ij})^2}{a^2}
\Big)
+\frac{1}{8}
\big(2\gamma_{ik}\gamma_{jl}-\gamma_{ij}\gamma_{kl}\big)
\frac{\partial_k\partial_l\gamma_{ij}}{a^2}
\bigg]\,.
\end{align}
Here we dropped the fluctuations of the lapse and shift, i.e. took the
decoupling limit, because their contributions to bispectra are higher
order in the slow-roll parameter $\epsilon=-\dot{H}/H^2$ or the
couplings $M_i$'s and $\bar{M}_i$'s.  Also note that the relation
between the Goldstone boson $\pi$ and the scalar perturbation $\zeta$ is
given by $\zeta=-H\pi$ at the linear order.  Similarly, the $\delta
K_\mu^\nu\delta K_\nu^\mu$ term is transformed in the decoupling limit
as
\begin{align}
\nonumber
&\int d^4x\sqrt{-g}\,\frac{-\bar{M}_3^2}{2}\delta K_\mu^\nu\delta K_\nu^\mu\\
\nonumber
&\to
\int d^4x\,a^3\frac{-\bar{M}_3^2}{2}\bigg[
\frac{(\partial^2\pi)^2}{a^4}+\frac{1}{4}\dot{\gamma}_{ij}^2
-\frac{1}{2}\dot{\gamma}_{ij}\frac{\partial_k\gamma_{ij}\partial_k\pi}{a^2}\\
\nonumber
&
-2\gamma_{ij}\frac{\partial_i\partial_j\pi\partial_k^2\pi}{a^4}
-\frac{1}{2}\Big(\ddot{\gamma}_{ij}+H\dot{\gamma}_{ij}-\frac{\partial_k^2\gamma_{ij}}{a^2}\Big)\frac{\partial_i\pi\partial_j\pi}{a^2}
\\
\label{KK_pi}
&
+\dot{\pi}\frac{4\partial^2\partial_i\pi\partial_i\pi-2(\partial_i\partial_j\pi)^2+4(\partial^2\pi)^2}{a^4}+\ldots\bigg]\,,
\end{align}
where the dots stand for higher order terms in perturbations and terms
proportional to $\dot{\bar{M}}_3$.  We notice that (\ref{KK_pi})
contains scalar-tensor-tensor type cubic interactions as well as
scalar-scalar-tensor and scalar-scalar-scalar type interactions.  In
Table~\ref{table}, we summarize what types of interactions arise
in the decoupling limit
from the operators in~(\ref{EFT_action}).  There, we find that the
scalar-tensor-tensor interaction, the $\gamma^2\pi$-type interaction,
arises only from the operator $\delta K_\mu^\nu\delta K_\nu^\mu$, while
$\gamma\pi^2$- and $\pi^3$-type interactions arise also from other
operators.  In this sense, we could say that the scalar-tensor-tensor
bispectrum is sensitive to the tensor sound speed $c_\gamma$ and is
enhanced by the deformation of $c_\gamma$.

\begin{table}
\begin{center}
\scalebox{0.85}{
\begin{tabular}{|c||c|c|c|c|c||c|c|c|c|}
\hline
&&&&&&&&&
\\[-2.5mm]
operator &
\,\,$\dot{\pi}^2$\,\, &
$\displaystyle\frac{(\partial_i\pi)^2}{a^2}$ &
$\displaystyle\frac{(\partial_i^2\pi)^2}{a^4}$ &
\,$(\dot{\gamma}_{ij})^2$\, &
$\displaystyle\frac{(\partial_k\gamma_{ij})^2}{a^2}$ &
\,\,$\gamma^3$\,\, &
\,$\gamma^2\pi$\, &
\,$\gamma\pi^2$\, &
\,\,$\pi^3$ \,\,\\[2.7mm]
\hline
&&&&&&&&&
\\[-4mm]
$S_0$ &$\checkmark$&$\checkmark$&& $\checkmark$ &
$\checkmark$ &  $\checkmark$&  &$\checkmark$&  \\
&&&&&&&&&
\\[-4mm]
\hline
&&&&&&&&&
\\[-4mm]
$(\delta g^{00})^2$ &
$\checkmark$&&&&
 &  &&  &$\checkmark$\\
&&&&&&&&&
\\[-4mm]
\hline
&&&&&&&&&
\\[-4mm]
$\delta g^{00}\delta K$ &
&$\checkmark$&&&&
 &  &  $\checkmark$&$\checkmark$\\
&&&&&&&&&
\\[-4mm]
\hline
&&&&&&&&&
\\[-4mm]
$(\delta K)^2$ &
&&$\checkmark$&&
 &  & & $\checkmark$&$\checkmark$\\
&&&&&&&&&
\\[-4mm]
\hline
&&&&&&&&&
\\[-4mm]
$\delta K_\mu^\nu\delta K_\nu^\mu$ &
&&$\checkmark$&$\checkmark$&&
 &  $\checkmark$&  $\checkmark$&$\checkmark$\\[0.5mm]
\hline
\end{tabular}}
\caption{Operators relevant to dispersion relations of primordial perturbations and the induced cubic interactions in the decoupling limit.}
\label{table}
\end{center}
\end{table}

\subsection{Evaluation of scalar-tensor-tensor bispectrum}

\begin{figure}[t]
\begin{center}
\includegraphics[width=60mm]{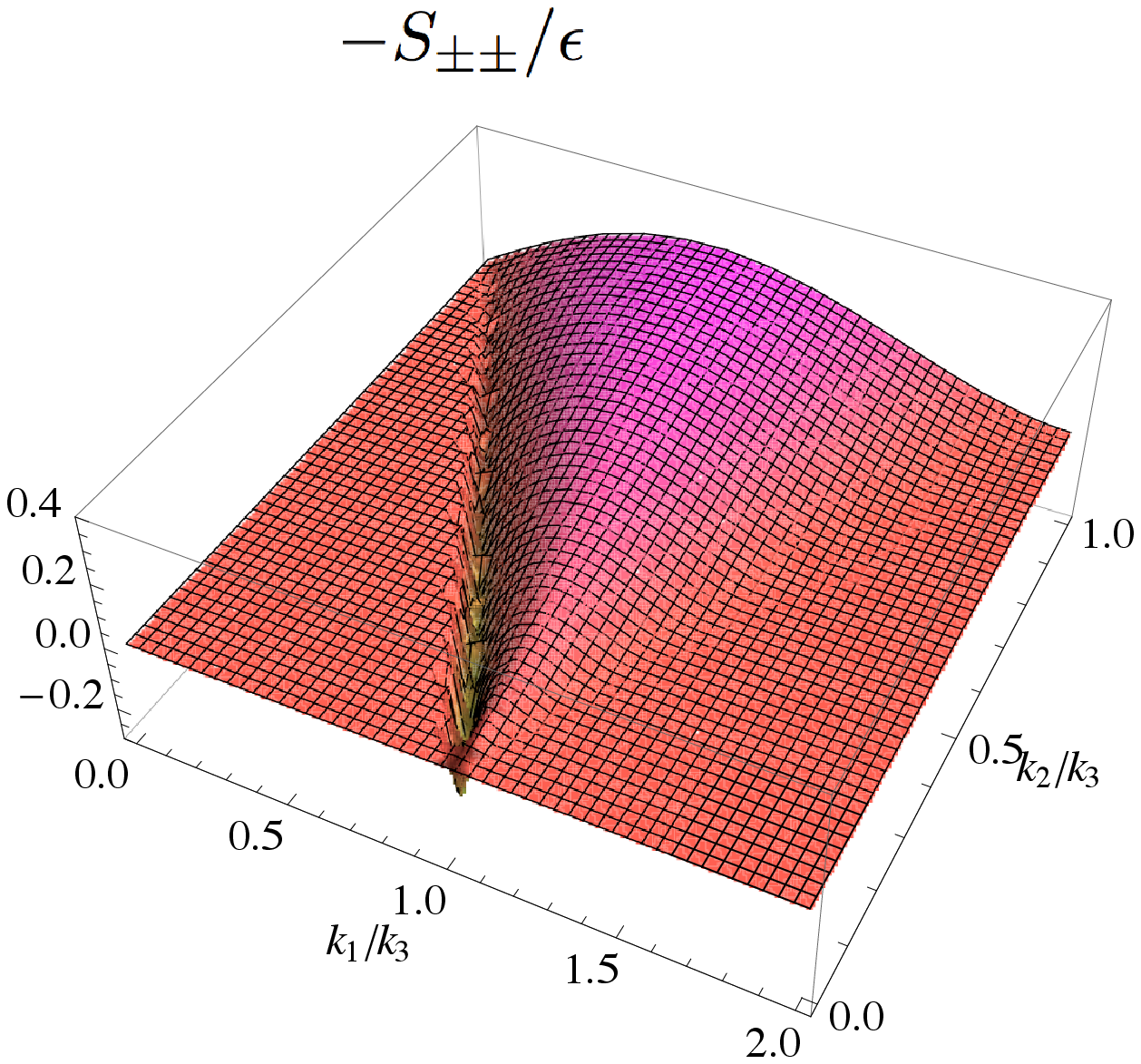}
\end{center}
\vspace{-5mm}
\caption{Shape function $S_{\pm\pm}(k_1,k_2,k_3)$ for $c_\gamma=0.8$}
\label{S_pmpm}
\begin{center}
\includegraphics[width=60mm]{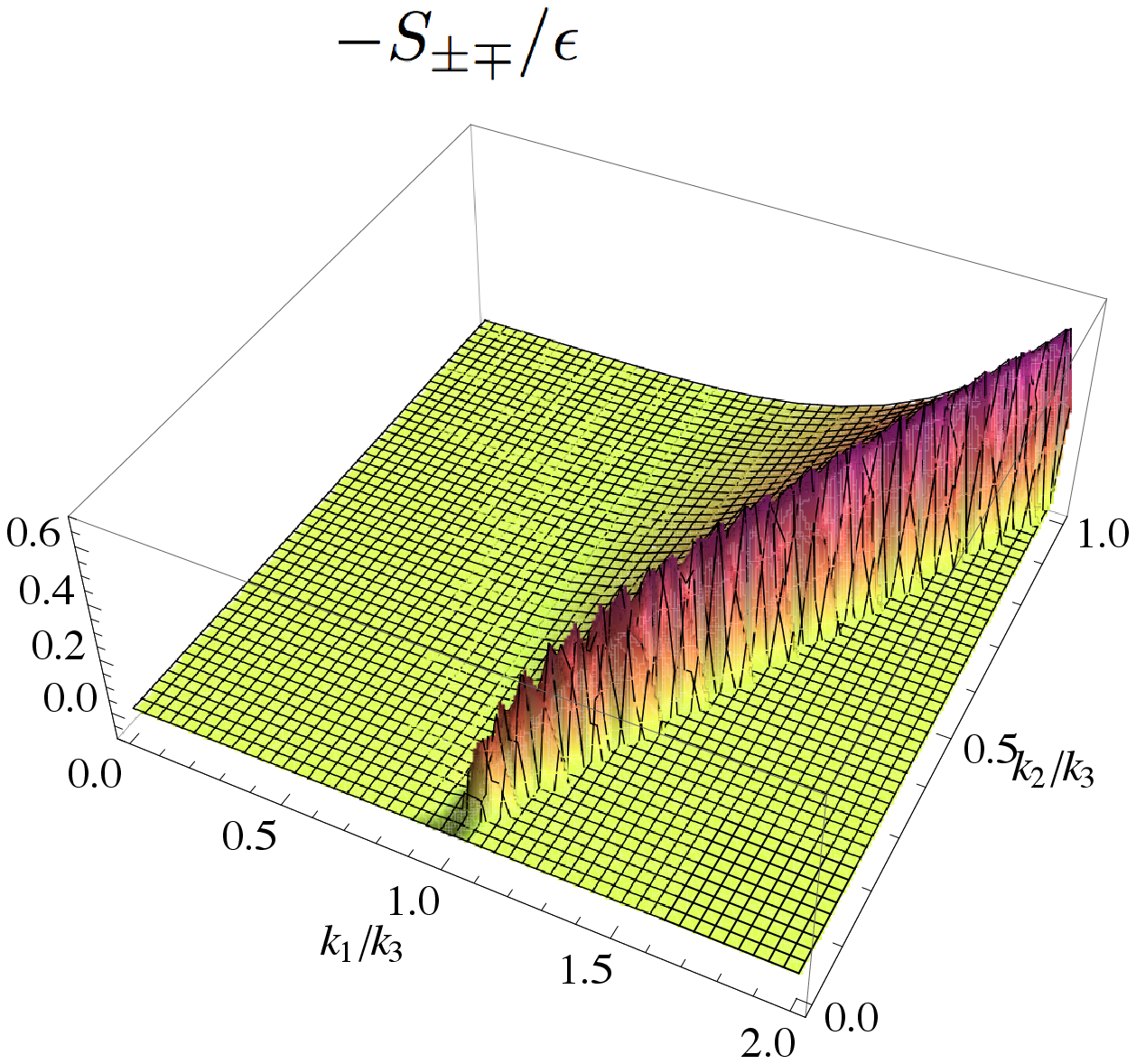}
\end{center}
\vspace{-5mm}
\caption{Shape function $S_{\pm\mp}(k_1,k_2,k_3)$ for $c_\gamma=0.8$}
\label{S_pmmp}
\end{figure}

We then take a closer look at the scalar-tensor-tensor cross
correlations by a concrete in-in formalism computation.  By using the
relation $\zeta=-H\pi$, the scalar-tensor-tensor correlation can be
expressed in terms of the Goldstone boson $\pi$ as
\begin{align}
\nonumber
\langle\zeta_{\mathbf{k}_1}\gamma_{\mathbf{k}_2}^{s_2}\gamma_{\mathbf{k}_3}^{s_3}\rangle
&=
-H\langle\pi_{\mathbf{k}_1}\gamma_{\mathbf{k}_2}^{s_2}\gamma_{\mathbf{k}_3}^{s_3}\rangle\,,
\end{align}
whose source is
the following interaction in (\ref{KK_pi}):
\begin{align}
\int d^4x\,a^3\frac{\bar{M}_3^2}{4}\dot{\gamma}_{ij}\frac{\partial_k\gamma_{ij}\partial_k\pi}{a^2}\,.
\end{align}
As given in Table~\ref{table}, the kinetic term of $\pi$ can be
modified by various correction terms.  However, for simplicity, let us
take the free theory action for $\pi$ as
\begin{align}
-M_{\rm Pl}^2\int d^4x\,a^3
\dot{H}\Big(
\dot{\pi}^2-\frac{(\partial_i\pi)^2}{a^2}
\Big)\,.
\end{align}
The Goldstone boson $\pi$ is then quantized as
\begin{align}
\pi&=\int\frac{d^3k}{(2\pi)^3}
\left[
a_{\bf k}u_{k}(t)+a_{\bf -k}^\dagger u_{k}^*(t)
\right]
e^{i{\bf k}\cdot {\bf x}}\,,
\end{align}
with the standard commutation relation
\begin{align}
[a_{{\bf k}},a^\dagger_{{\bf k}'}]=(2\pi)^3\delta({\bf k}-{\bf k}^\prime)
\end{align}
and the Bunch-Davies mode function
\begin{align}
u_k=\frac{1}{2M_{\rm Pl}\epsilon^{1/2}k^{3/2}}
(1+ik\tau)e^{-ik\tau}\,.
\end{align}
Introducing the shape function $S_{s_2,s_3}(k_1,k_2,k_3)$ for the
scalar-tensor-tensor bispectrum (normalized by the scalar powerspectrum) as
\begin{align}
\nonumber
&\langle\zeta_{\mathbf{k}_1}\gamma_{\mathbf{k}_2}^{s_2}\gamma_{\mathbf{k}_3}^{s_3}\rangle
\\
&=
(2\pi)^3\delta^3\Big(\sum_i\mathbf{k}_i\Big)
\frac{(2\pi)^4\mathcal{P}_\zeta^2}{k_1^2k_2^2k_3^2}\,
S_{s_2,s_3}(k_1,k_2,k_3)\,,
\end{align}
we can easily calculate it by the in-in formalism as
\begin{align}
S_{s_2,s_3}(k_1,k_2,k_3)
&=
\epsilon_{ij}^{s_2}({\bf k}_2)\epsilon_{ij}^{s_3}({\bf k}_3)\widetilde{S}(k_1,k_2,k_3)
\,.
\end{align}
Here $\widetilde{S}$ is a helicity-independent part given by
\begin{align}
\nonumber
&\widetilde{S}(k_1,k_2,k_3)
\\
\nonumber
&=
\epsilon\left(c_\gamma^{-2}-1\right)\frac{({\bf k}_1\cdot {\bf k}_2)k_3^2}{2k_1k_2k_3}
\left(\frac{1}{K}+\frac{k_1+c_\gamma k_2}{K^2}+\frac{2k_1k_2}{K^3}\right)\\
&\quad
+({\bf k}_2\leftrightarrow {\bf k}_3)\,,
\end{align}
where $K=k_1+c_\gamma k_2+c_\gamma k_3$ and we used
$\bar{M}_3^2=(1-c_\gamma^{-2})M_{\rm Pl}^2$.  An explicit form of the
helicity part is
\begin{align}
\label{hel_part}
\epsilon_{ij}^{s_2}({\bf k}_2)\epsilon_{ij}^{s_3}({\bf k}_3)
&=
\left\{\begin{array}{ccc}
\frac{1}{2}(1-\cos\theta)^2 & \,\,{\rm for} \,\,& s_2=s_3 \,,
\\
\frac{1}{2}(1+\cos\theta)^2 &\,\, {\rm for}\,\,& s_2\neq s_3\,,
\end{array}\right.
\end{align}
where $\theta$ is the angle between the momenta ${\bf k_2}$ and ${\bf
k_3}$.  For $s_2=s_3$, the helicity part (\ref{hel_part}) takes its
maximum value $2$ when ${\bf k_2}$ and ${\bf k_3}$ are
antiparallel and vanishes when they are parallel.  For
$s_2=- s_3$, vice versa.  The total shape function $S_{s_2,s_3}$ can
then be classified into $S_{\pm\pm}$ and $S_{\pm\mp}$.  As depicted in
Fig.~\ref{S_pmpm}, the peak of $S_{\pm\pm}$ is around
$(k_1,k_2,k_3)\simeq(c_\gamma k,k,k)$, where the three modes have the
same sound horizon size.
The $c_\gamma$-dependence at this point is given by
\begin{align}
S_{\pm,\pm}(c_\gamma k,k,k)
&=-\frac{(1-c_\gamma^2)(4-c_\gamma^2)^2(2+15c_\gamma)}{432c_\gamma^{3}}\epsilon\,.
\end{align}
On the other hand, as shown in Fig.~\ref{S_pmmp}, $S_{\pm,\mp}$
has a peak at $(k_1,k_2,k_3)=(2k,k,k)$, where ${\bf k}_2$ and ${\bf
k}_3$ are parallel and the helicity part (\ref{hel_part}) is maximized.
The $c_\gamma$-dependence of the peak size is given by
\begin{align}
S_{\pm,\mp}(2k,k,k)
&=-\frac{(1-c_\gamma)(6+7c_\gamma+3c_\gamma^2)}{2c_\gamma^2(1+c_\gamma)^2}\epsilon\,.
\end{align}
It should be noted that, when the tensor sound speed is unity,
the scalar-tensor-tensor bispectrum vanishes in the decoupling limit.
Then,
the leading contribution
comes from fluctuations of the lapse and shift,
and becomes
$\mathcal{O}(\epsilon^2)$, which is higher order in the slow-roll parameter $\epsilon$
compared to the case with $c_\gamma\neq1$.
The scalar-tensor-tensor bispectrum is then enhanced as 
\begin{align}
\frac{\langle \zeta \gamma\gamma\rangle_{c_\gamma\neq1}}{\langle \zeta \gamma\gamma\rangle_{c_\gamma=1}}\sim \mathcal{O}\Big(\frac{c_\gamma^{-2}-1}{\epsilon}\Big)\,,
\end{align}
when the tensor sound speed is modified.
Because of the factor $1/\epsilon$,
a significant enhancement can occur
even if the deformation of the tensor sound speed is not so large, say $c_\gamma\sim0.8$.
Therefore,
the enhancement of the scalar-tensor-tensor bispectrum can be  a powerful probe
for the modified tensor sound speed.

\subsection{On scalar bispectrum}

Finally, we make a brief comment on scalar bispectra induced by the
modified tensor sound speed.  As in (\ref{KK_pi}), the $\delta
K_\mu^\nu\delta K_\nu^\mu$ operator induces cubic interactions of $\pi$,
which can be a source of large scalar non-Gaussianities.  Indeed, if
$\delta K_\mu^\nu\delta K_\nu^\mu$ is the only relevant operator and
other operators do not come into the game, the scalar nonlinearity
parameter $f_{\rm NL}$ can be estimated as
\begin{align}
f_{\rm NL}\sim \mathcal{O}\big((c_\gamma^{-2}-1)/\epsilon\big)\,.
\end{align}
Since it is enhanced by the inverse of the slow-roll parameter, one may
think that the current null observation of scalar non-Gaussianities can
constrain the tensor sound speed as $c_\gamma^{-2}-1\lesssim \epsilon$.
However, as shown in Table~\ref{table}, various operators can induce
$\pi^3$-type interactions and the scalar non-Gaussianities can be easily
reduced in the presence of other operators.
For example,
when the $\delta K^2$ operator is relevant
and $\bar{M}_2^2=-\bar{M}_3^2$,
\begin{align}
S=S_0-\frac{\bar{M}_3^2}{2}\int d^4x\sqrt{-g}\big(\delta K_\mu^\nu\delta K^\mu_\nu
-\delta K^2\big)\,,
\end{align}
the cubic interactions of $\pi$ exactly cancel out in the decoupling limit.\footnote{
If we go beyond the decoupling limit,
the leading contributions to the scalar bispectrum
may arise from terms with fluctuations of the lapse and shift,
which are higher order in $c_\gamma^{-2}-1$ or $\epsilon$.
While such contributions are negligible as long as $c_\gamma^{-2}-1$ is small,
they become relevant when $c_\gamma\lesssim\frac{1}{\sqrt{2}}$
and more careful discussions are required in such a parameter region.
See~\cite{full} for details.}
In fact, the generalized Galileon
\cite{Deffayet:2011gz,Horndeski:1974wa} accommodates this type of
combination in the action \cite{Gleyzes:2013ooa}. Thus, additional
symmetries or tunings may decrease the scalar
non-Gaussianities.  In this way, scalar bispectra depend on various
operators and the information of the tensor sound speed is obscured.  In
contrast, the scalar-tensor-tensor bispectra are more sensitive to the
tensor sound speed and can be a powerful tool to measure it.

\subsection{Conclusion}

In this {\it Letter}, we investigated the relation between
non-Gaussianities of primordial perturbations and the sound speed of
tensor perturbations, based on the EFT approach.  We found that the
modification of the tensor sound speed induces a significant enhancement
of the scalar-tensor-tensor cross bispectrum, rather than the tensor
auto-bispectrum. This situation is in sharp contrast with the case
of the curvature perturbations, in which their auto-bispectra are
significantly enhanced by their reduced sound velocity.  When the sound
speed of tensor perturbations is reduced, the scalar-tensor-tensor
bispectrum is enhanced by a factor of $(c_\gamma^{-2}-1)/\epsilon$
compared to the case of $c_\gamma=1$ and such an enhancement makes
it easy to detect the CMB bispectra of two B-modes and one temperature
(or one E-mode) anisotropies especially. Thus, the scalar-tensor-tensor bispectrum
and its relevant bispectra of the CMB can be powerful probes for the
reduced tensor sound speed and the gravitational structure for
inflation.
\\

\paragraph*{Acknowledgments}

The work of T.N. is supported in part by the Special Postdoctoral
Researcher Program at RIKEN. The work of M.Y. is supported in part by
the JSPS Grant-in-Aid for Scientific Research on Innovative Areas
No.~24111706 and the JSPS Grant-in-Aid for Scientific Research
No.~25287054.


\end{document}